\documentstyle[amssymb,12pt]{article}

\newif\iffigs\figstrue
\figsfalse

\iffigs
  \input epsf
\else
\fi

\batchmode
  \newfont{\footscrfont}{rsfs10}
  \newfont{\footbbbfont}{msbm10}
\errorstopmode

\newif\ifscrf\scrftrue
\ifx\footscrfont\nullfont
  \scrffalse
\fi

\newif\ifamsf\amsftrue
\ifx\footbbbfont\nullfont
  \amsffalse
\fi

\def\ppnumber{\vbox{\baselineskip14pt\hbox{CK-TH-2000-001, }
\hbox{hep-th/0001062}}}
\def\ppdate{January 2000}
\def\pplogo{\vbox{\kern-\headheight\kern -15pt
\halign{##&##\hfil\cr&{
\ppnumber}\cr\rule{0pt}{2.5ex}&\ppdate\cr}
}}
\makeatletter
\date{}
\def\dedicatory#1{\def\@date{\normalsize\it#1}}
\def\subjclass#1{\def\@thefnmark{}\@footnotetext{1991
    {\it Mathematics Subject Classification.} #1}}
\def\keywords#1{\def\@thefnmark{}\@footnotetext{
    {\it Key words and phrases.} #1}}

\def\ps@firstpage{\ps@empty \def\@oddhead{\hss\pplogo}%
  \let\@evenhead\@oddhead 
}
\def\maketitle{\par
 \begingroup
 \def\thefootnote{\fnsymbol{footnote}}
 \def\@makefnmark{\hbox
 to 0pt{$^{\@thefnmark}$\hss}}
 \if@twocolumn
 \twocolumn[\@maketitle]
 \else \newpage
 \global\@topnum\z@ \@maketitle \fi\thispagestyle{firstpage}\@thanks
 \endgroup
 \setcounter{footnote}{0}
 \let\maketitle\relax
 \let\@maketitle\relax
 \gdef\@thanks{}\gdef\@author{}\gdef\@title{}\let\thanks\relax}

\def\abstract{\if@twocolumn
\section*{{\bf Abstract}}
\else \small
\begin{center}
{\bf ABSTRACT}
\end{center}
\quotation
\fi}

\def\thebibliography#1{\section*{References\@mkboth
 {REFERENCES}{REFERENCES}}\small\list
 {[\arabic{enumi}]}{\settowidth\labelwidth{[#1]}\leftmargin\labelwidth
 \advance\leftmargin\labelsep
 \usecounter{enumi}}
 \def\newblock{\hskip .11em plus .33em minus .07em}
 \sloppy\clubpenalty4000\widowpenalty4000
 \sfcode`\.=1000\relax}

\newif\iffn\fnfalse

\@ifundefined{reset@font}{\let\reset@font\empty}{} 
\long\def\@footnotetext#1{\insert\footins{\reset@font\footnotesize
    \interlinepenalty\interfootnotelinepenalty
    \splittopskip\footnotesep
    \splitmaxdepth \dp\strutbox \floatingpenalty \@MM
    \hsize\columnwidth \@parboxrestore
   \edef\@currentlabel{\csname p@footnote\endcsname\@thefnmark}\@makefntext
    {\rule{\z@}{\footnotesep}\ignorespaces
      \fntrue#1\fnfalse\strut}}}

\makeatother

\begin{document}
\setcounter{page}0
\title{\LARGE {\bf Gauge and Gravitational Couplings 
from Modular Orbits in Orbifold
Compactifications}\\[10mm]}
\author{
{\bf C. Kokorelis}\\[0.1cm]
\normalsize CITY University Business School\\
\normalsize Department of Investment, Risk Management and Insurance\\
\normalsize Frobischer Crescent, Barbican Centre,
London, EC2Y 8HB,U.K\\[5mm]
}

{\hfuzz=10cm\maketitle}

\def\Large{\large}
\def\LARGE{\large\bf}

\vskip 1cm

\begin{abstract}
We discuss
the appearance of modular functions at the one-loop gauge and
gra-
\newline
vitational couplings in (0,2) non-decomposable N=1 four dimensional orbifold
compactifications
of the heterotic string.
 We define the limits for the existence of states 
causing singularities
in the moduli space in the perturbative regime for a generic vacuum of
the heterotic string. The "proof" provides evidence for the explanation
of the stringy Higgs effect.

\end{abstract}

\newpage

\section{\bf Introduction}

The purpose of this paper is to examine the appearance of one-loop
threshold
corrections in gauge and gravitational
couplings, in four dimensional non-decomposable orbifolds of the
heterotic string.
In 4${\cal D}$ $N=1$ orbifold compactifications
the process of integrating out 
massive string modes, causes
the perturbative one-loop threshold corrections\footnote{which
receive non-zero moduli dependent
corrections from the $N=2$ unrotated complex planes},
to receive non-zero corrections in the form
of automorphic functions of the target space modular group.
The one-loop threshold corrections
can be calculated either by calculation of string amplitudes
or by the sum over modular orbits.
The latter technique will be used in this work.

At special points in the moduli space, previously massive states become
massless,
and contribute to gauge symmetry enhancement. The net result  of the
appearance of massless
states in the running gauge coupling constants appears in the form
of a dominant logarithmic term.
In section two we will discuss the logarithmic term effect and 
suggest that its
appearance, due to the
nature of the underlying modular integration, sets specific limits in the
mass of the previously massive states that becoming massless at the enhanced
symmetry point.

In addition, in this paper, we are particularly interested
in the calculation of one-loop threshold effects in non-decomposable
orbifolds, using the technique
of summing over modular orbits, that arise after integrating out
the moduli dependent
contributions of the heavy string modes. The last technique have been used
in a variery of contexts, such as,  
the calculation of target space free energies of
toroidal compactifications in \cite{oova} and of Calabi-Yau compactifcation
models \cite{feko}, in addition to the calculation of
threshold effects to gauge and
gravitational couplings in $N=1$ 4${\cal D}$
decomposable orbifold compactifications \cite{lust1} and the
calculation of target space free energies and $\mu$-term contributions
in $N=1$ 4${\cal D}$
non-decomposable orbifold compactifications in \cite{kokos2}.
In section three we will discuss the appearance of automorphic functions
of ${{\Gamma}_o(3)}_{T,U}$ via the calculation of
modular orbits of
target space free energies and thus the threshold corrections, generalizing
to non-decomposable orbifolds the discussion in \cite{lust1} for
decomposable ones.
In sections four and five we will complete the picture by extending the
calculation of one-loop threshold effects to gauge and gravitational
couplings respectively,
using the sum over modular orbits (SMO),
to $N=1$ 4${\cal D}$ non-
decomposable orbifolds.
We will exhibit the application of SMO by examiming
a $Z_6$ $N=1$ non-decomposable orbifold which exhibits
a
${\Gamma^o(3)}_T \times {\Gamma^o(3)}_U$ target space duality group
in one of its two dimensional untwisted subspaces.
The gauge embedding in the gauge degrees of freedom will not be specified,
apart for its $T^2$ torus subspace part,
and kept generic in order for the threshold effects
to be dependent only on the Wilson line context of its two
dimensional subspace.

\section{Massless singularity limit}

In general, if one wants to describe globally the moduli space
and not just the small field deformations
of an effective theory around a specific vacuum solution, one
has to take into account the number of massive states
that become massless at a generic point in moduli space.  
This is a necessary, since the full duality group   
$SO(22,6;Z)_T$ mixes massless with massive modes\cite{lust2}.
It happens because there are transformations of O(6,22,Z) acting as
automorphisms of the Lorentzian lattice metric of
$\Gamma^{(6,22)}=\Gamma^{(6,6)} \oplus \Gamma^{(0,16)}$
that transform massless states into massive
states.

Let us consider now the $T_2$ torus, coming from the decomposition
of the $T_6$ orbifold into the form $T_2 \oplus T_4$.  
At the large radius limit of the $T^2$
it was noticed \cite{cfilq} that in the presence of states that
become massless at a point in moduli space
e.g, when the  $T \rightarrow U$, the threshold corrections
to the gauge coupling constants receive
the most dominant logarithmic contribution in the form,
\begin{equation}
{\triangle}_a(T,{\bar T})\;\; {\approx}\;\; b_{a}^{\prime} \;\;
\int_{\Gamma} {\frac{d^2 \tau}{\tau_{2}}} e^{-M^2(T){\tau_{2}}}
\;\;{\approx} - b_a^{\prime} {\log{M^2\left( T \right)}},
\label{koko1}
\end{equation}
where $b_ {a}^{\prime}$ is the contribution to the $\beta$-function
from the
states that become massless at the point $T=U$.
Strictly speaking the situation is sightly different.
We will argue that if
we want to include in the string effective field theory large field
deformations and to describe the string Higgs effect \cite{higgs} 
and not only small field fluctuations, eqn.(\ref{koko1}) must be modified.
We will see that massive states which become massless at specific
points in the moduli space do so, only if the values of the untwisted
moduli dependent masses are
between certain limits.
In \cite{cfilq} this point was not emphasized and it was presented
in a way
that the appearance of the singularity in eqn. (\ref{koko1})
had a general validity for
generic values of the mass parameter.

We introduce the function Exponential Integral $E_1(z)$
\begin{equation}
E_1 (z) = \int_z^\infty \frac{e^{-t}}{t} dt\;\;\;\;\;(|arg\;z|)<\pi),
\label{tion11}
\end{equation}
with the expansion
\begin{equation}
E_1 (z)= - \gamma - lnz - \sum_{n=1}^\infty \frac{(-)^n z^n}{nn!}.
\label{tion12}
\end{equation}  
It can be checked that for values of the parameter $|z|>1$, the $lnz$ term
is not the most dominant, while for $0<|z|<1$ it is.

In the latter case \cite{abraste} the $E_1(z)$ term is
approximated\footnote{
The "Exponential Integral" $E_1(x)$ for $0 \le x \le 1$ is $E_1(x)= -ln(x) +
\alpha_{0} + \alpha_{1}x + \alpha_{2}x + \alpha_{3}x +\alpha_{4}x +
\alpha_{5}x + \epsilon(x),
|\epsilon(x)| \le 2 \times 10^{-7}$,
with the numerical constants $a_i$ to be given by 
\begin{eqnarray}
\alpha_{0} = - 5.77&\alpha_{1} = 0.99&\alpha_{2}=-0.25  \nonumber \\
\alpha_{3} = -0.55&\alpha_{4} = -0.009&\alpha_{5}=0.00107
\end{eqnarray}}
as
\begin{equation}
E_1 (z) = -ln(z) +a_{0} +a_{1} z +a_{2} z^2 +a_{3} z^3 +a_{4} z^4 +
a_{5} z^5 + \epsilon(z).
\label{tion14}
\end{equation}
Take now the form of eqn.(\ref{koko1}) explicitly
\begin{equation}
\triangle(z,{\bar z}) \approx
b^{\prime}_a \int_{|\tau_1|<{1/2}}
d\tau_1 \int_{\sqrt{1 - \tau_1^2}}^{\infty}
e^{-M^2(T)\tau_2} .
\label{lop12}
\end{equation}
Then by using eqn.(\ref{tion11}) in
eqn.(\ref{lop12}),
we can see that the $- b^{\prime}_a\;ln M^2(T)$ indeed arise.
Notice now, that the limits of the integration variable $\tau_1$
in the world-sheet integral in eqn.(\ref{koko1}) are between $-1/2$
 and $1/2$.  
Then especially for the value $|1/2|$ the lower limit in the
integration
variable $\tau_2$ takes its lowest value e.g
$(1-\tau_1^2)^{1/2} =(1 -(1/2)^2)^{1/2} = \sqrt{3}/2$.
Use now eqn.(\ref{tion12}).
Rescaling the $\tau_2$ variable in the integral,
and using the condition $0<z<1$
we get the
necessary condition for the logarithmic behaviour to be
dominant\footnote{Restoring units in the Regge slope
parameter $a^{\prime}$.}
\begin{equation}
0<M^2(T)<\frac{4}{{\sqrt3}a^{\prime}}.
\label{wqe12}
\end{equation}
This means that the
dominant behaviour of the threshold corrections appears in the form
of a logarithmic singularity, only when the moduli scalars  
satisfy the above limit.

We know that for particular values of the moduli
scalars, the low energy effective theory appears to have
singularities, which are due to the appearance of charged massless
states in the physical spectrum.
At this stage, the contribution of the mass to the low
energy gauge coupling parameters is given by \cite{cfilq}
\begin{equation}
M^2 \rightarrow  - n_H|T-p|^{2},
\label{asas}
\end{equation}
where the $n_{H}$ represents the number of states $\phi_{H}$ which
become massless at the point $p$.

The parameter $(a^{\prime}\sqrt{3}/{4}) M^2$ must always be between the
limits
zero and one in order that the dominant contribution of the physical
singularity to $\triangle$ to be in the "mild" logarithmic
form (\ref{asas}).
Therefore, the complete picture of the threshold effects,
when the asymptotic behaviour of the threshold corrections
is involved, reads
\begin{equation}
\frac{1}{g_a^2(\mu)}=\frac{k_a}{g_{string}^2} + \\
\frac{b_a}{16\pi^2} \ln \frac {M_{string}^2}{\mu^2} -\\
\Theta(- M^2 + \frac{4}{{\sqrt 3}a'}) b^{\prime}_{a} \log M^2(T),
\label{gatos1}
\end{equation}
where $\Theta$ is the step function.
Threshold effect dependence on the $\Theta$
function, takes place in Yang - Mills theories, via the
decoupling theorem \cite{algaa}. The contribution of the
various thresholds decouples from the full theory, and the
net effect
is the appearance of mass suppressed corrections to the 
physical quantities.
Their direct effect on the low energy effective theory is
the appearance of the automorphic functions of the
moduli dependent masses, after the integration of    
the massive modes.

So far, we have seen that the theory can always approach the
enhanced symmetry point behaviour from a general massive point on the
moduli space under specific conditions.
For "large" values of the moduli masses the enhanced symmetry point
can be approached if its mass is inside the limit (\ref{wqe12}). 
 Remember that at the point $T=p$
eqn.(\ref{gatos1}) breaks down, since at this point perturbation theory
is not valid any longer.


\section{\cal  Target space automorphic functions from string
compactifications}

Before looking at the appearance of
automorphic functions in the one-loop gauge and gravitational couplings 
of 4$\cal D$ orbifold compactifications,
using the sum over modular orbits,
 we need some background
on the mass operator moduli dependence in orbifolds.
For orbifold compactifications, where the underlying internal torus does not
decompose into a $T_6 = T_2 \oplus T_4 $ , the $Z_2 $ twist associated with
the reflection $- I_2 $ does not put any additional constraints on the
moduli $U$ and $T$. As a consequence the moduli space of the untwisted
subspace is the same as in toroidal compactifications and orbifold sectors
which have the lattice twist acting as a $Z_2$, give non-zero threshold
one-loop corrections to the gauge coupling constants in $N=1$ supersymmetric
orbifold compactifications.

In the study of the untwisted moduli space, we will assume initially that
under the action of the internal twist there is a sublattice of the Narain
lattice ${\Gamma}_{22,6}$ in the form ${\Gamma}_{22,6} \supset \Gamma_{2} \oplus
\Gamma_{4}$ with the twist acting as $-\;I_2$ on $\Gamma_{2}$. In the
general case, we assume that there is always a sublattice\footnote{%
this does not correspond to a decomposition of the Narain
lattice as ${\Gamma%
}_{22,6} = {\Gamma}_{q+2,2} \oplus \dots$ since the gauge lattice $%
\Gamma_{16}$ is an Euclidean even self-dual lattice. So the only way for it
to factorize as $\Gamma_{16} = {\Gamma_{q}} \oplus {\Gamma_r}$, with $%
\;q+r=16$, is when $q=r=8$.} ${\Gamma}_{q+2,2} \oplus {\Gamma}_{r+4,4}
\subset {\Gamma}_{16,6}$, where the twist acts as $-\;I_{q+4}$,
on ${\Gamma}_{q+2,2}$ and with eigenvalues different than -I on ${\Gamma}
_{r+4,4}$.
In this case, the mass formula for the untwisted subspace ${\Gamma}_{q+2,2}$
depends on the factorised form $P_{R}^{2} = v^{T} {\phi}{\phi^{T}}v$, with $%
v^{T}$ taking values as a row vector, namely as
\begin{equation}
v^{T}=(a^{1},\dots,a^{q};n^{1},n^{2};m_{1},m_{2}).  \label{kln1}
\end{equation}
The quantities in the parenthesis represent the lattice coordinates of the
untwisted sublattice ${\Gamma}_{q+2,2}$, with $a^{1}, \dots, a^{q}$ the Wilson
line quantum numbers and $n^{1},n^{2},m_{1},m_{2}$ the winding and momentum
quantum numbers of the two dimensional subspaces.

Let us consider first the generic case of an orbifold where the internal
torus factorizes into the orthogonal sum 
$T_6 = T_2 \oplus T_4$ with the $Z_2$ twist acting on the 2-dimensional torus
lattice. We will be interested in the mass formula of the untwisted subspace
associated with the $T_2$ torus lattice. We consider as before that there is
a sublattice of the Euclidean self-dual lattice ${\Gamma}_{22,6}$ as ${\Gamma%
}_{q+2,2} \oplus {\Gamma}_{20-q\;;\;4} \subset {\Gamma}_{22,6}$. In this case,
the momentum operator factorises into the orthogonal components of the
sublattices with $(p_{L};p_{R}) \subset {\Gamma}_{q+2;2}$ and $(P_{L};P_{R})
\subset {\Gamma}_{20-q\;,\;4}$. As a result the mass operator 
factorises into the form
\begin{equation}
\frac{\alpha^{\prime}}{2} M^{2} = p_{R}^{2} + P_{R}^{2} + 2N_{R}.
\label{arxidi1}
\end{equation}
On the other hand, the spin operator S for the ${\Gamma}_{q+2;2}$ sublattice
changes as
\begin{equation}
p_{L}^{2} - p_{R}^{2}= 2(N_{R} + 1 - N_{L}) + P_{R}^{2} - P_{L}^{2}=2 n^{T}
m + q^{T} C q,  \label{arxidi2}
\end{equation}
where C is the Cartran metric operator for the invariant directions of the
sublattice ${\Gamma}_{q}$ of the $\Gamma_{16}$ even self-dual lattice.
In eqn's (\ref{arxidi1},\ref{arxidi2}), we discussed the level
matching condition in the case of a $T_6$ orbifold admitting an orthogonal
decomposition.

Let us now consider the gauge symmetry enhancement\footnote{
The information about the nature of singularities
will then used in the
 calculation of the modular orbits.}
 of the $Z_6-II-b$ orbifold.
 This orbifold is defined on the torus
lattice $SU(6) \times SU(2)$ and the twist in the complex basis is defined
as $\Theta = exp((2,-3, 1) \frac{2 \pi i}{6})$.
This orbifold is non-decomposable in the sense that the action of the
lattice twist does not decompose into the orthogonal sum $T_6 = T_2 \oplus T_4 
$ with the fixed plane lying in $T_2$. The orbifold twists $\Theta^2$ and $%
\Theta^4$, leave the third and complex plane unrotated . The lattice in
which the twists $\Theta^2$ and $\Theta^4$ act as a lattice automorphism is
the $SO(8)$. In addition there is a fixed plane which lies in the $SU(3)$
lattice and is associated with the $\Theta^3$ twist.

Consider now the k-twisted sector of a six-dimensional orbifold of the the
heterotic string associated with a twist $\theta^{k}$. The twisted sector
quantum numbers have to satisfy 
\begin{eqnarray}
Q^{k}n=n,\;\;Q^{*k}m=m,\;\;M^{k}l=l,  \label{axid1}
\end{eqnarray}
where Q defines the action of the twist on the internal lattice and M
defines the action of the gauge twist on the $E_{8} \times E_{8}$ lattice.

In the $Z_{6}-II-b$ orbifold, for the $N=2$ sector associated to the
$\Theta^2$ twist, $n^{T}m=m_{1}n^{1}+3m_{2}n^{2}$, and
\begin{equation}
m^2 = \sum_{{m_1, m_2 }{n^1,n^2}} \frac{1}{Y} |-T
U^{\prime}n^2+iTn^1-iU^{\prime}m_1+3m_2|^2_{U^{\prime}= U+2} = {\cal M}/{(Y/2)},
\label{kavou}
\end{equation}
with
$Y = (T +{\bar T})(U+{\bar U})$.
The quantity $Y$ is associated with the K\"{a}hler potential, $K = - \log Y$.
The target space duality group is found to be $\Gamma^0(3)_T \times
\Gamma^0(3)_{U^{\prime}}$, where $U^{\prime}= U +2$.
Mixing of the equations (\ref{arxidi1}, {\ref{arxidi2}) gives us the
following equation
\begin{equation}
p_{L}^{2} - \frac{\alpha^{\prime}}{2}M^{2} = 2(1-N_{L}-\frac{1}{2}
P_{L}^{2}) = 2 n^{T} m + q^{T} C q . \label{oki1}
\end{equation}
The previous equation gives us a number of different orbits invariant
under $SO(q+2,2;Z)$ transformations :

a) the untwisted orbit with $2 n^{T} m + q^{t} C q = 2$.
In this orbit, $N_{L}=0,\;P_{L}^{2}=0 $. In particular, when $M^{2}=0$,
this orbit is
associated with the string Higgs effect. The string Higgs effect appears
as a special solution of the (\ref{oki1}) at the point where
$p_{L}^{2}=2$, where additional massless particles may appear.

b) the untwisted orbit where
$2 n^{T} m + q^{t} C q = 0$
 where, $2 N_{L}+P_{L}^{2} = 2 $. This is the orbit relevant to the
calculation of threshold corrections to the gauge couplings, without
taking into account the enhanced gauge symmetry points.

c) The massive untwisted orbit with
$2N_{L} +P_{L}^{2} \geq 4  $. Now always $M^{2} \geq 0 $.
This orbit will be of no use to our attempt of exhibiting
the singular behaviour of threshold
corrections.
\newline
Let us now consider,
for the orbifold $Z_{6}-II-b$, the modular orbit associated with
the string Higgs effect. We are looking for points in the moduli
space where singularities associated with the additional massless particles
appear and have as a result gauge group enhancement. This point correspond
to $T=U$ with $m^2 = n^2 = 0 $ and $m^1 = n^1 = \pm 1$. At this point the
gauge symmetry is enhanced to $SU(2) \times U(1)$. In particular, the left
moving momentum for the two dimensional untwisted subspace yields
\begin{equation}
p_{L}^{2}= \frac{1}{2T_{2}U_{2}} |{\bar T}Un_{2} - {\bar T} n_{1} -i
U^{\prime}m_{1} + 3 m_{2}|^{2}=2,  \label{oki3}
\end{equation}
while
\begin{equation}
p_{R}^{2} = \frac{1}{2T_{2}U^{\prime}_{2}}|-{T}U^{\prime}n^{2} + i{T} n^{1}
-iU^{\prime}m_{1} + 3 m_{2}|^{2} = 0.  \label{oki2}
\end{equation}
At the fixed point of the modular group $\Gamma^{o}(3)$, $\frac{\sqrt{3}}{2}%
(1 + i \sqrt{3})$, there are no additional massless states, so there is no
further enhancement of the gauge symmetry.

We will now use eqn.(\ref{kavou}) to calculate the stringy one-loop
threshold corrections to the gauge coupling constants coming from the
integration of the massive compactification modes with $(m,m^{\prime},n,n^{%
\prime}) \neq (0,0,0,0)$. The total contribution to the threshold
corrections, coming from modular orbits and associated with the presence of
massless particles, is connected to the existence of the following%
\footnote{%
,we calculate only ${\sum} {\log {\cal M}}$ since  $\sum \log {\cal M}%
^{\dagger}$ is its complex conjugate,}
orbits\cite{feko,lust1,kokos2},
\begin{eqnarray}
{\Delta}_0&=&{\sum}_{2 n^t m + {q^T {\cal C} q} =2 } {\log {\cal M}}|_{reg} 
\nonumber \\
{\Delta}_1 &=&{\sum}_{2 n^t m+{q^T{\cal C}q}=0}{\log {\cal M}}|_{reg}.
\label{polop}
\end{eqnarray}
In the previous expressions, a regularization procedure is assumed that
takes place, which renders the final expressions finite, as infinite sums
are included in their definitions.
 Morover, we demand that the regularization procedure for $%
e^{\Delta} $ has to respect both modular invariance and holomorphicity.
The regularization is responsible for the
subtraction of a moduli
independent quantity from the infinite sum e.g ${\sum}_{n\;,m \in orbit} {%
\log {\cal M}}$.
The regularization procedure for the case of a decomposable orbifold,
where the threshold corrections are invariant under the $SL(2,Z)$,
were discussed in \cite{feko}. The general case of the regularization
procedure for
the case of non-decomposable orbifolds,
where the threshold corrections are invariant under subgroups of $SL(2,Z)$,
 was discussed
in \cite{kokos2}.

Let us consider first the orbit relevant for the string Higgs effect . This
orbit is associated with the quantity $2 n^T m + q^T {\cal C} q = 2$,
 where $n^T m = m_1 n^1 + 3 m_2 n^2$.
The total contribution from the previously mentioned orbit yields :
\begin{equation}
{\Delta}_{0}{\propto}{\sum}_{n^T m + q^2 =1} \log {\cal M}  \nonumber \\
= {\sum}_{n^T m =1 , q=0} \log {\cal M} + \sum_{n^T m =0 , q^2 = 1} \log 
{\cal M} + \sum_{n^T m = -1 , q^2 = 2} \log {\cal M} + \dots  \label{wzerooo}
\end{equation}
We must notice here that we have written the sum \cite{lust1} over the states
associated with the $SO(4,2) $ invariant orbit $2 n^T m + q^T {\cal C} q = 2 
$ in terms of a sum over $\Gamma^0(3)$ invariant orbits $n^T m = constant$ .
We will be first considering the contribution from the orbit $2 n^T m + q^T 
{\cal C} q = 0 $. Note that we are working in analogy with calculations associated
with topological free energy considerations \cite{oova,feko,lust1,kokos2}.
From the second
equation in eqn.(\ref{polop}), considering in general the $S0(4, 2)$ coset,
we get for example that
\begin{eqnarray}
\Delta_{1} \propto \sum_{n^T m + q^2 = 0} \log {\cal M} = \sum_{n^T m =0,
q= 0 }\log {\cal M} + \sum_{n^T m =-1,q^2 = 1}\log {\cal M} + \dots
\label{wzerooooa}
\end{eqnarray}

Consider in the beginning the term $\sum_{n^T m =0 , q = 0 } {\log {\cal M}} 
$. We are summing up initially the orbit with $n^T m = 0; (n, m) \neq (0,0)$,
\begin{equation}
{\cal M} = 3 m_2 - im_1 U^{\prime}+ in^1 T + n^2 (-U^{\prime}T + B C)+ {%
q\;dependent\;terms}.  \label{poli}
\end{equation}
We calculate the sum over the modular orbit $n^T m + q^2 = 0$. As in \cite
{lust1} we calculate initially the sum over massive compactification states
with $q_1= q_2= 0$ and $(n, m) \neq (0,0)$. Namely, the orbit
\begin{eqnarray}
\sum_{n^T m =0,\; q = 0} \log {\cal M}= \sum_{(n,m)\neq (0,0)}\log(3m_2
-im_1 U^{\prime}+in_1 T+n_2 (-U^{\prime}T)) &  \nonumber \\
+\;BC \sum_{(n,m)\neq (0,0)}{\frac{n_2}{(3m_2 -im_1 U^{\prime}+in_1 T-n_2
U^{\prime}T)}} +{\cal O}((BC)^2).  \label{polia}
\end{eqnarray}
The sum in relation (\ref{polia}) is topological (it excludes oscillator
excitations) and is subject to the constraint $3m_2 n^2 + m_1 n^1 = 0$.
Its
solution receives contributions from the following sets of integers:
\begin{equation}
m_2 = r_1 r_2 \;,\;n_2 = s_1 s_2\;,\;m_1 = - 3 r_2 s_1\;,\;n_1 = r_1 s_2,
\label{polib}
\end{equation}
\begin{equation}
m_2 = r_1 r_2 \;,\;n_2 = s_1 s_2\;,\;m_1 = - r_2 s_1\;,\;n_1 = 3 r_1 s_2,
\label{polibb}
\end{equation}
and
\begin{eqnarray}
\sum_{n^T m=0,\; q = 0} \log {\cal M} = \log[\left({\eta^{-2}(T)} {\frac{1}{3%
}}{\eta^{-2}{(\frac{U^{\prime}}{3})}}\right)(1- 4\;BC\; (\partial_T \log {%
\eta(T)})&\times&  \nonumber \\
(\partial_U^{\prime}\log {\eta (\frac{U^{\prime}}{3}}))]\;\; + \log [\;(({%
\eta^{-2}(U^{\prime})}{\frac{1}{3}}){\eta^{-2}(\frac{T}{3})}) (1-
4BC(\partial_T \log &\times&  \nonumber \\
{\eta(\frac{T}{3})})(\partial_U^{\prime}\log\eta(U^{\prime})))\;]\; + {\cal O%
}((BC)^2).&&  \nonumber \\
\label{fufutos}
\end{eqnarray}
The previous expression is associated with the
non-perturbative \cite{feko,lust1,kokos2}
gaugino generated superpotential ${\cal W}$, which comes
 by direct integration of
the string massive orbifold modes.
The contribution of this term could give rise to a direct Higgs
mass in the effective action and represents a particular solution to the $%
\mu $ term problem. These issues are discussed in \cite{kokos2}.
The threshold contribution of (\ref{fufutos}) to the modular orbit ${\Delta}_1$ of
eqn. (\ref{polop})
is obtained by substituting (\ref{fufutos}) in (\ref{wzerooooa}) yielding
\begin{eqnarray}
{\Delta}_1 \propto \log[\left({\eta^{-2}(T)} {\frac{1}{3%
}}{\eta^{-2}{(\frac{U^{\prime}}{3})}}\right)(1- 4\;BC\; (\partial_T \log {%
\eta(T)})&\times&  \nonumber \\
(\partial_U^{\prime}\log {\eta (\frac{U^{\prime}}{3}}))]\;\; + \log [\;(({%
\eta^{-2}(U^{\prime})}{\frac{1}{3}}){\eta^{-2}(\frac{T}{3})}) (1-
4BC(\partial_T \log &\times&  \nonumber \\
{\eta(\frac{T}{3})})(\partial_U^{\prime}\log\eta(U^{\prime})))\;]\; + \dots
.&&  
\label{fufutostos}
\end{eqnarray}
The previous discussion was restricted to small values of the 
Wilson lines where our $(0,2)$ orbifold goes into a $(2,2)$.
We turn now our discussion to the contribution from the first equation in
(\ref{polop}) which is relevant to the stringy Higgs effect. Take for example
the expansion (\ref{wzerooo}). Let's examine the first orbit corresponding to
${\Delta}_{0,0} =\sum_{n^T m =1, q=0} \log {\cal M}$. This orbit is
the one for which some of the previously massive states, now become
massless. At these points ${\Delta}_{0,0}$ have to exhibit the
logarithmic singularity. In principle we could predict, in the simplest case
when the Wilson lines have been switched off that ${\Delta}_{0,0}$ may be
given by  
\begin{eqnarray}
\Delta_{0,0} = \sum_{n^T m =1} \log (T U^{\prime}n^2 + Tn^1- U^{\prime}m_1 +
3m_2 ) 
= \log \{(\omega(T) - \omega(U^{\prime}))^{\xi} \times \nonumber\\
\{ \eta(T)^{-2}\; \eta(\frac{U^{\prime}}{3})^{-2}
+\;\eta(\frac{T}{3})^{-2}\;\eta(U^{\prime})^{-2}\} +\dots, \label{repjj}
\end{eqnarray}
where $\omega(T)$ is the hauptmodul \cite{ko} for the subgroup
${\Gamma^o(3)}_T$, namely $\omega =  [\eta(T/3)/{\eta}(T)]^{12}$. 
The behaviour of $\Delta_0$ term reflects the\footnote{%
in the following we will be using the variable $U$ instead of $U^{\prime}$.}
fact that at the points with $T=U$, generally previously massive states
becoming massless, while the eta-terms are needed for consistency under
modular transformations. Finally, the integers $\chi$, $\zeta$ have to be
calculated from a string loop calculation or by directly performing the sum. 
Note that for the R.H.S of (\ref{repjj}) there is no known way
of directly performing the sum.

After this parenthesis, we continue our discussion by turning on, Wilson
lines. When we turn the Wilson lines on, for the $SO(4,2)$ orbit of the
relevant untwisted two dimensional subspace, $\Delta_{0,0}$ becomes 
\begin{equation}
\Delta_{0,0} = \sum_{n^T m =1} \log \{ 3m_2 - i m_1 U + i n_1 T - n_2 ( U T
- B C )\}.  \label{delzero}
\end{equation}
The sum after using an ansatz, similar to \cite{lust1}, and keeping
only lowest order terms satisfy
\begin{eqnarray}
\Delta_{0,0} & = & \log \left( \omega(T) - \omega (U) - B C \;X
(T,U) \right)^{\xi} + \log \{\eta(T)^{-2} \;\eta(\frac{U}{3})^{-2}
+  \nonumber \\
&+&\;\eta(\frac{T}{3})^{-2} \;\eta(U)^{-2} -  B C \;{\cal Y}(T,U)\}
+ \dots \nonumber\\
\label{w00bcc}
\end{eqnarray}
The functions $X(T,U)$,$\;$${\cal Y} (T,U)$, may be
calculated by the demand of duality invariance. 
Let us first discuss the calculation of $X (T,U)$.
Demanding duality invariance of the first term in (\ref{repjj}),
under ${\Gamma^o(3)}_U$ modular transformations,
we get that $X (T,U) $ has to obey - to the lowest non-trivial order in B C - the
transformation
\begin{equation}
X(T,U) \stackrel{{\Gamma^o(3)}_U}{\rightarrow} (i \gamma U + \delta )^2 \; X(T,U) - i \gamma ( i \gamma
U + \delta) \; (\partial_T \omega(T)).
\label{asdwae11}
\end{equation}
In (\ref{asdwae11}) we have used the fact that 
under
the ${\Gamma^o(3)}_U$ target space duality transformations
\begin{eqnarray}
U & \stackrel{{\Gamma^o(3)}_U}{\rightarrow}  & \frac{\alpha U -i \beta}{i \gamma U + \delta} \;,\;\;\; T
\rightarrow T - i \gamma \frac{BC}{i \gamma U + \delta} \;,\;\;\;
\alpha \delta - \beta \gamma = 1,  \nonumber \\
B &\rightarrow& \frac{B}{i \gamma U + \delta} ,\;\;\;\; C \rightarrow \frac{C%
}{i \gamma U + \delta},\;\;\;\beta = 0\;mod\;3,  \label{utbctransf}
\end{eqnarray}
which leave the
tree level K\"ahler potential
\begin{equation}
K = - \log[(T+ {\bar T})(U + {\bar U})- ({\bar B}+ C)( B + {\bar C})]
\label{opli12}
\end{equation}
invariant \cite{ANTO},
the following transformation is valid
\begin{equation}
\omega(T) - \omega(U) \stackrel{{\Gamma^o(3)}_U}{\rightarrow} \;\; \omega(T) - \omega(U) \; -
\; i
\gamma \; \frac{BC}{i \gamma U
+ \delta} \;\; (\partial_T \omega(T)).
\label{asdwea1}
\end{equation}
In a similar way invariance under ${\Gamma^o(3)}_T$ transformations
\begin{eqnarray}
T & \stackrel{{\Gamma^o(3)}_T}{\rightarrow}  & \frac{\alpha T -i \beta}{i \gamma T + \delta} \;,\;\;\; U
\rightarrow U - i \gamma \frac{BC}{i \gamma T + \delta} \;,\;\;\;
\alpha \delta - \beta \gamma = 1,  \nonumber \\
B &\rightarrow& \frac{B}{i \gamma T + \delta} ,\;\;\;\; C \rightarrow \frac{C%
}{i \gamma T + \delta},\;\;\;\beta = 0\;mod\;3,  \label{utbctransfa}
\end{eqnarray}
which leave (\ref{opli12}) invariant, $X(T,U)$ has to transform as 
\begin{equation}
X(T,U) \stackrel{{\Gamma^o(3)}_T}{\rightarrow} (i \gamma T + \delta )^2 \; X(T,U) + i \gamma ( i \gamma
T + \delta) \; (\partial_U \omega(U)) ,
\end{equation}
up to the lowest order in B C . So far we have described the properties of
$X(T,U)$ under modular transformations. The final form of our function,
which has to respect the proper modular transformations, and to reveal the
presence of physical singularities in the quantum moduli space reads
\begin{eqnarray}
X(T,U) = -3 \partial_U \{ \log \eta^2(\frac{U}{3})\} \;
\omega^{\prime}(T) + \partial_T \{ \log \eta^2(T) \} \;
\omega^{\prime}(U)& +  \nonumber \\
\beta \{\omega (T) - \omega(U) \} \{ \eta^4(T) \eta^4(\frac{U}{3})\} +
{\cal O}((BC)^2),
\label{termos1}
\end{eqnarray}
where $\beta$ is a constant which may be decided from a loop calculation.
Lets us now try to determine the Y-term in 
\begin{equation}
{\cal D}=
\log \left(\eta(T)^{-2} \;\eta(\frac{U}{3})^{-2} + \eta(\frac{T}{3})^{-2}
\;\eta(U)^{-2}-
B C 
\;{\cal Y} (T,U)\right)
\label{sdeawq1}
\end{equation}
of (\ref{w00bcc}). It should transform with modular weight
-1 under ${\Gamma^o(3)}_U$ transformations. In this case
we find that $\cal Y$ has to transform, up to order BC as
\begin{eqnarray}
{\cal Y}(T,U) \stackrel{{\Gamma^o(3)}_U}{\rightarrow} (i \gamma U + \delta ) \;{\cal Y}(T,U)
- i \gamma \{\eta^{-2}(\frac{U}{3}) (\partial_T \;\eta^{-2}(T)) +
(\partial_{\frac{T}{3}} \eta^{-2}(\frac{T}{3}))  \;\eta^{-2}(U)\}. 
\label{prop1}
\end{eqnarray}
On the other hand, if we demand that it transforms with modular weight -1
under ${\Gamma^o(3)}_T$ we get that, up to lowest order in BC, 
\begin{equation}
{\cal Y}(T,U) \stackrel{{\Gamma^o(3)}_T}{\rightarrow} (i \gamma T + \delta )
\; {\cal Y}(T,U) - i \gamma
\{ (\partial_{\frac{U}{3}} \eta^{-2}(\frac{U}{3}) \;\eta^{-2}(T) +
 (\partial_U \eta^{-2}(U))  \;\eta^{-2}(\frac{T}{3})) \}. 
 \label{prop2}
\end{equation}
The modular properties (\ref{prop1}), (\ref{prop2}) and the
presence of the physical
singularities in our moduli space fix the function ${\cal Y}(T, U)$ up
to order $(BC)^2$
as
\begin{eqnarray}
{\cal Y}(T,U) = \{
\eta^{-2}(T) \;\eta^{-2}(\frac{U}{3})(\partial_T \;\eta^2(T))
(\partial_{\frac{U}{3}}\; \eta^2 (\frac{U}{3})) + \nonumber\\
\eta^{-2}(U) \;\eta^{-2}(\frac{T}{3}) 
(\partial_{\frac{T}{3}} \;\eta^2(\frac{T}{3}))
(\partial_{U}\; \eta^2 (U)) \}
+\;\rho [\;(\eta^2(T) \; \eta^2(\frac{U}{3})) +
\eta^2(\frac{T}{3})  \eta^2(U) ], \nonumber\\
\label{termos1}
\end{eqnarray}
where $\rho$ may be decided from a a loop calculation.
It follows now, from (\ref{w00bcc}) that
 $e^{\triangle_{0,0}}$,   reads
up to the order $(BC)^2$,
\begin{eqnarray}
e^{\triangle_{0,0}} \propto [(\omega(T)-\omega(U))^{\xi}
\;\{\eta(T)^{-2} \;\eta(\frac{U}{3})^{-2} +
\;\eta(\frac{T}{3})^{-2} \;\eta(U)^{-2}\}
- BC {\cal Y} [\omega(T)-\omega(U)]^{\xi} - \nonumber\\
-\xi\;(\omega(T)-\omega(U))^{\xi - 1}\; B C X
\{\eta(T)^{-2} \;\eta(\frac{U}{3})^{-2} +
\;\eta(\frac{T}{3})^{-2} \;\eta(U)^{-2}\}
 + {\cal O}((BC)^2).
 \label{asdxzc12}
\end{eqnarray}
We must notice here that the expression for $e^{\triangle_{0,0}}$
transforms with modular weight -1 under the ${\Gamma^o(3)}_{U, T}$ modular
 transformations (\ref{utbctransf},\ref{utbctransfa}).
This is natural since from the relations \cite{feko},
\begin{eqnarray}
Z=e^{- F_{fermionic}} = - det(( \frac{{\cal M}^{\dagger}}{Y^{\dagger}})
\frac{ {\cal M}}{Y}) = -
\frac{|{\cal W}|^2}{Y},\nonumber\\
F_{fermionic} = \sum_{(n,m) \neq (0,0)} \log
det((\frac{{\cal M}^{\dagger}}{Y^{\dagger}} )\frac{ {\cal M}}{Y}  ),
\label{free}
\end{eqnarray}
where $F_{fermionic}$ the fermionic free energy,
the quantity $e^{\triangle_{0,0}}$ is identified
with $\cal W$, the superpotential.

%

\section{Threshold corrections to gauge couplings}

We will now analyze the threshold corrections to the gauge couplings,
due to the
integration of massive modes, in the case
of $N=1$ symmetric $(2,2)$ non-decomposable orbifold compactifications
of the heterotic string.
When considering an effective locally supersymmetric theory, we have
to distinguish between the kind of renormalized physical couplings involved
in the theory. These are the cut-off dependent Wilsonian gauge couplings and
the moduli and momentum dependent effective gauge couplings (EGC).
Let us consider contributions to the EGC from the $(2,2)$
symmetric non-decomposable $Z_6 -II-b$ orbifold considered in the previous
section. We want to examine the EGC when the embedding
in the gauge degrees of freedom is such that the gauge group in the
"observable" sector gets broken to a subgroup by turning on Wilson line moduli fields
B, C on the untwisted subspace of the non-decomposable orbifold.
We consider a general embedding in the gauge degrees of freedom such
that the gauge group, in the "hidden" sector remains unbroken, namely
$E_8^{\prime}$. The contributions to the EGC receive contributions
from all the $N=2$ sectors of the nondecomposable orbifold. Here for
simplicity reasons we will consider only the contribution to the
thresholds of the
EGC from the $N=2$ $\Theta^2$ sector that were examined in the previous
section.
 We examine first the contributions to the EGC from the
unbroken $E_8^{\prime}$ gauge group.
In this case the threshold corrections $\triangle_{E_8^{\prime}}$
receive contributions from the
untwisted $N=2$ orbit, $2 n^T m + q^T C q = 0 $, of (\ref{fufutos}) yielding 
\begin{eqnarray}
\triangle_{E_8^{\prime}} =
c({E_8^{\prime}})
\log\left(9 |\eta(T) \; \;\eta(\frac{U}{3})|^4 |1 - BC
(\partial_T \log\eta^2(T))(\partial_U \log \eta^2(U)|^{-2} \right)+\nonumber\\
+\; c({E_8^{\prime}})
\log\left(9 |\eta(U)  \;\eta(\frac{T}{3})|^4 |1 - BC
(\partial_T \log\eta^2(\frac{T}{3}))(\partial_U \log \eta^2(U)|^{-2} \right).
\label{fufutos1}
\end{eqnarray}
The full threshold corrections to the EGC receive an additional
contribution from the massless modes, due to K\"ahler and sigma model
anomalies equal to
\begin{equation}
\triangle_{massive} =  C_a K - 2\sum_r T_a(r) \log det g_r, 
\label{fufutos2}
\end{equation}
where $C_a= -C(G_a) + \sum_r T_a(r)$, $C_a$ the quadratic Casimir of the
gauge group $G_a$, K is the K\"ahler potential of the $N=2$ unrotated subspace,
 the sum is over the chiral matter superfields transforming in a representation
 r of $G_a$ and $g_r$ is the $\sigma$-model metric of the massless sector
 that the
matter fields in the representation r belong.
The equation for the EGC associated to the $E_8^{\prime}$ for a
scale $p^2 << M_{E_8^{\prime}}$, after taking
into account (\ref{fufutos2}) and the contribution
fron the massive
states that have been
integrated out, namely eqn. (\ref{fufutos1}), becomes
\begin{eqnarray}
\frac{1}{g^2_{E_8^{\prime}}} = \frac{S + {\bar S}}{2} +
\frac{b_{E_8^{\prime}}}{16 \pi^2} \log \frac{M_{string}^2}{p^2} -
{\tilde a}_{E_8^{\prime}} \log\left( (T + {\bar T})(B + {\bar B}) - ({\bar B} + C)
({\bar C} + B)\right)
 \nonumber\\
+\;c({E_8^{\prime}})
\log\left(9 |\eta(T) \eta(\frac{U}{3})|^4 |1 - BC
(\partial_T \log\eta^2(T))(\partial_U \log \eta^2(\frac{U}{3})|^{-2}
\right)+\nonumber\\
+\; c({E_8^{\prime}})
\log\left(9 |\eta(U)  \;\eta(\frac{T}{3})|^4 |1 - BC
(\partial_T \log\eta^2(\frac{T}{3}))(\partial_U \log \eta^2(U)|^{-2} \right),
\label{fufutos3}
\end{eqnarray}
where $b_{E_8^{\prime}}= -3 c({E_8^{\prime}})$ and
\begin{equation}
{\tilde a}_{E_8^{\prime}} = C_{E_8^{\prime}}= - c({E_8^{\prime}}).
\label{aaq123}
\end{equation}

In the following we will consider that, in the running of
EGC at the "observable" sector,
beyond the high energy string
scale there is an additional scale, e.g $M_I$, for which supersymmetry remains
unbroken and the gauge group G, sitting at the high energy scale, gets
spontaneously broken at a subgroup.
By inspection of (\ref{repjj}) we can realize that below the string scale
$M_{string}$ there is an additional scale given by
$M_I = |\omega(T)- \omega(U)|M_{string}$.
This is exactly the scale corresponding to gauge symmetry enhancement
to $SU(2) \times U(1)$.
Lets us now try to calculate the running gauge coupling for
the two additional massless vector multiplets\footnote{
In the case of $N=1$ four dimensional
compactifications of heterotic string vacua, the moduli of the invariant
subspace belongs to vector multiplets.}
 present in the spectrum at the
point $T=U$.
Note that
the running of the gauge couplings for points different than $T=U$,
between the scales $M_I$ and $ M_{string}$,
is given by
\begin{equation}
\frac{1}{g^2(M_I^{2})} = \frac{1}{g^2(M_{string}^2)} +
\frac{b_a}{16 \pi^2} \log \frac{M_{string}^2}{M_I^2} + \triangle_{massive},
\label{fufutos4}
\end{equation}
where $\triangle_{massive}$ is given in (\ref{fufutos2}).
Here, $b_a =-3 c(G_a) + \sum_{C} T_a(r_C)  - 3 \sum_V T_a(r_V) $, with the first
sum runs over the chiral
matter superfields transforming under a representation $r_C$ of the gauge
group
with $T_a(r_C)= Tr_{r_C}(T^2_a)$, the second sum runs over light vector
multiplet representations $r_V$, and $T_a$ denotes a generator of the
gauge group.
In the
 case that the gauge coupling of
the vector multiplets is in the region 
$p^2 << M_{I}$, we get
\begin{equation}
\frac{1}{g^2_{U(1)}(p^2)} = \frac{1}{g^2 (M_I^2)}
+ \frac{{\tilde b}_a}{16 \pi^2}\log\frac{M_I^2}{p^2} + \triangle,
\label{fufutos6}
\end{equation}
where
\begin{equation}
\triangle = - \frac{a_{U(1)} }{16 \pi^2} \{\log\left(
9|\eta(T) \eta(\frac{U}{3})|^4 \right)
+ log\left(9|\eta(U) \eta(\frac{T}{3})|^4 \right)\}.
\label{fufutos55}
\end{equation}
Here,
\begin{equation}
a_{U(1)} = - c(U(1)) +\sum_C T_{U(1)}( 1 + 2 n_{U(1)}),
\label{adzc123}
\end{equation}
where $n_{U(1)}$ the modular weights of the light chiral superfields.
Note that  
the moduli metric of the untwisted $N=2$ plane, from (\ref{fufutos2}),
$g_r = ((T + {\bar T})(U + {\bar U }))^{n_C}$,
 with $n_C$ the modular weight
of the light chiral superfields.
Let us now apply (\ref{fufutos4}, \ref{fufutos6}, \ref{fufutos55}, \ref{adzc123})
to the running gauge coupling
belonging to the 2 additional vector mupliplets present
in the spectrum above the threshold scale $M_I$, for the $Z_6 -II-b$ orbifold,
\begin{eqnarray}
&{\frac{1}{g_{U(1)}(p^2)} =\frac{S + {\bar S}}{2} + {\frac{\hat{b}_{U(1)}}{%
16 \pi^2}} \log {\frac{M^2_{string}}{p^2}} + {\frac{({\hat{b}}_{U(1)} -
b_{U(1)})}{16 {\pi}^2}}} {\log{({\omega(T)} - {\omega(U)} )}^{2}}-{\frac{%
a_{U(1)}}{16{\pi}^{2}}}  \nonumber \\
&\{\log \left( (T+{\bar T})(U +{\bar U})9|\eta(\frac{U}{3}) \eta(T)
|^4 \right) + \log\left((T+{\bar T})(U +{\bar U})9|\eta(U) \eta(%
\frac{T}{3})|^{4}\right)\}.
\label{ret}
\end{eqnarray}
Here, ${\tilde{b}_{U(1)}} =0$, since $c_{U(1)}=0$ and there are no
hypermultiplets charged under the $U(1)$. In the same way, 
$a_{U(1)}=0$, since the gauge group under
the additional threshold scale $M_I$ is abelian. The coefficient $b_{U(1)} $
equals ${\hat b}_{U(1)} + 2 b_{vec}^{N=2}$, where $b_{vec}$ the contribution
from the $\beta$-function coefficients of the $N=2$ vector multiplets
which are massless above the threshold scale and 2 counts their multiplicity.
The additional threshold scale
beyond the traditional string tree level unification scale is the one
associated with the term $\omega(T) - \omega(U)$. The threshold scale 
 is associated
with the enhancement of the abelian part of the gauge group to $SU(2)$.

\section{Threshold corrections to gravitational couplings}

Let us now discsuss contributions to the
running gravitational couplings in $(2,2)$ symmetric $Z_N$ orbifold
constructions of the heterotic string.

For $(0,2)$$\;$ $Z_{N}$ orbifolds the effective low energy action of the
heterotic string is 
\begin{equation}
{\cal L} = {\frac{1}{2}}{\cal R} + {\frac{1}{4}}{\frac{1}{g_{grav}}\; {\cal C}
+ {\frac{1}{4}}{S_R}}\;(GB) + {\frac{1}{4}}{S_I}\;R_{abcd}R^{abcd},
\end{equation}
where ${S_R} \equiv (S + {\bar S})$, $S_I \equiv 2ImS $. We have used the
conventional choice for the gravitational couplings is $1/{g_{grav}}\;
\equiv \;{S_R}$ , while $GB$ is the Gauss-Bonnet combination 
\begin{equation}
4\;(GB) = {{\cal C}^{2}} - 2 {\cal R}_{ab}^{2} +{\frac{2}{3}}{{\cal R}^{2}}
\end{equation}
and ${\cal C}$ the Weyl tensor ${\cal C}_{abcd}$. When the above relation
is written in the form 
\begin{equation}
{\cal L} \propto {{\triangle}^{grav}}(T,{\bar T})({\cal R}^{2}_{abcd} - 4 {\cal R}%
^{2}_{ab} + {\cal R}^{2}) +  \nonumber \\
{\Theta}^{grav}(T,{\bar T}) {\epsilon}^{abcd} {\cal R}_{abef} {\cal R}%
_{cd}^{ef},
\end{equation}
where $({\Theta}^{grav}(T,{\bar T})$ the CP-odd part of GB, then
the one-loop corrections \cite{anto2}, ${{\triangle}^{grav}}$,
to the gravitational action in $N=1$ decomposable orbifolds, in
the absence of Green-Schwarz mechanism, give
${{\triangle}^{grav}} \propto {
\tilde b}^{grav}_{N=2} \log(T + {\bar T})|\eta(iT)|^{4}$, where
${\tilde b}^{grav}_{N=2}$ the gravitational $\beta$-function coefficient
that receives non-zero contributions from the $N=2$ sectors. 

The corrections to the gravitational couplings considered up to know in the
literature, are concerned with the decomposable orbifolds. We will complete
the discussion 
of corrections to the running gravitational couplings by examining
non-decomposable orbifolds.
For the latter orbifolds the threshold corrections are expressed in terms of
automorphic functions belonging to subgroups of the inhomogeneous modular
group $PSL(2,Z)=SL(2,Z)/{\pm 1}$.

We focus our attention to the case of ${Z_6}-II-b$ orbifold. We
consider the case of vanishing Wilson lines in the $\Theta^2$ sector.
In the presence of the
threshold ${p^2} \ll {M_{I}^{2}} \ll {M_{string}^{2}}$, we get 
\begin{eqnarray}
{\frac{1}{{g_{grav}}^2 ({M_I}^2) }} = \frac{1}{g_{grav}^2(M_{string}^2)} +
 {\frac{b_{grav}}{16 \pi^2}}%
\log{\frac{M_{I}^2}{M_{string}^2}}
 - {\frac{a_{grav}^{\prime}}{16 \pi^2}} \log \left( {\eta^{4}(T)} {\eta^{4}(%
\frac{U}{3})}\; {9} \right) -
\nonumber\\
- {\frac{a_{grav}^{\prime}}{16 \pi^2}} \log \left(9 {\eta^{4}(%
\frac{T}{3})} {\eta^{4}(U)}\right)\;\;\;\;\;  \label{tyui1}
\end{eqnarray}
and
\begin{equation}
{\frac{1}{{g_{grav}}^2 (p^2) }} = {\frac{1}{{g_{grav}}^2 ({M_I}^2) }}
+ {\frac{{\tilde b}_{grav}}{16 \pi^2}} \log {\frac{M_{I}^2}{p^2}} 
- \frac{{\tilde a}_{grav}}{16 \pi^2} \log \left((T + {\bar T})(U + {\bar U})
\right).
\label{tyui2}
\end{equation}
Here $ \frac{1}{g_{grav}^2 (M_{string}^2)}={\frac{S + {\bar S}}{2}}$
the treel level coupling,
${\tilde b}_{grav}$, $b_{grav}$ the $\beta$-function coefficients  for the range $p^2 << M_I^2$,
$M_I^2 << p^2 << M_{string}$ respectively.
Note that in (\ref{tyui1}, \ref{tyui2}) we neglected the contributions for the $Z_6-II-b$
orbifold that are
coming from the other $N=2$ sectors.  For all our stydy and
conclusions
regarding  (\ref{tyui1}, \ref{tyui2}) we have considered that our orbifold
has only one $N=2$ sector, the $\Theta^2$ sector.
 If we want to consider
the full $Z_6 -II-b$ orbifold, we should add the holomophic contributions
from the other $N=2$ sector in addition to the contributions
to the $\beta$-function coefficients
of the fixed plane lying in the $SU(3)$ lattice, invariant under the
$\Theta^3$
twist, for which the contributions to the gravitational ruuning couplings
transform under $PSL(2,Z)$. The                                                      
${\tilde a}_{grav}$ comes from non-holomorphic contributions from
K\"{a}hler and ${\sigma}$-model anomalies and is given by
${\tilde a}_{grav} = \frac{1}{24} (21 + 1 - \dim G + \gamma_M +
{\sum}_{\tilde C}(1+2n_{\tilde C}))$,
where $\gamma_M$ is the contribution from modulinos.
 The ${\tilde a}_{grav}$ has been
calculated in the absence
of continuous Wilson lines \cite{anto2} as coefficients of the Gauss-Bonnet
term in the gravitational action and represents the contribution of the
completely rotated $N=2$ plane. In that case
${{\tilde a}_{grav}} = {\tilde b}_{grav}^{N=2}$.
The coefficient $a_{grav}^{\prime}$ has also been calculated in \cite{anto2}
and equals ${\tilde b}_{grav}^{N=2}$.
Moreover, because of the contribution of the
additional vector multiplet which become massless above the enhancement
scale $M_I$,
$(b_{grav} - {\tilde b}_{grav}) = \gamma_{grav}^C+ \gamma_{grav}^V$, with
$\gamma_{grav}^C$, $\gamma_{grav}^V$ 
the contributions to the gravitational $\beta$-function arising from the
decomposition of the additional $N=2$ vector multiplet in terms of its
 $N=1$ multiplets.
That happens
because any $N=2$ vector multiplet, can be decomposed into a
$N=1$ vector
multiplet and a $N=1$ chiral multiplet.
Substituting 
(\ref{tyui1}) into (\ref{tyui2}) we get 
\begin{eqnarray}
{\frac{1}{{g_{grav}}^2 (p^2) }} = {\frac{S + {\bar S}}{2}}
+ {\frac{{\hat b}_{grav}}{16 \pi^2}} \log {\frac{M_{string}^2}{p^2}}
- \frac{\gamma_{grav}^C+ \gamma_{grav}^V}{16 \pi^2}\log|\omega(T) -
\omega(U)|^2
\nonumber\\
 - {\frac{a_{grav}^{\prime}}{16 \pi^2}} \log((T + {\bar T})(U + {\bar U})
  {\eta^{4}(T)} {\eta^{4}(%
\frac{U}{3})9}) 
 - {\frac{a_{grav}^{\prime}}{16 \pi^2}} \log((T + {\bar T})(U + {\bar U})
  {\eta^{4}(U)} {\eta^{4}(
\frac{T}{3})9}),
\label{azxsaz1}
\end{eqnarray}
which is invariant under ${\Gamma_0(3})_{T,U}$ transformations.

Orbifolds, where the target space modular groups belong
to a subgroup of the modular group
may be found from further compactifying six-dimendional F-theory
compactifications on a
general Calabi-Yau 3-fold with an $F_n$ base where the order of the Mordell-Weyl
group may be three \cite{aspi,koko}.

\end{document}